RESEARCH ARTICLE

# DNA Linear Block Codes: Generation, Error-Detection, and Error-Correction of DNA Codeword

**Mandrita Mondal[1*] and Kumar S. Ray[2]**

[1]*Indian Statistical Institute, 203, B.T. Road, Kolkata 700108, India.*
[2]*Department of Computer Engineering & Applications, GLA University, Mathura, Uttar Pradesh, India.*

## Abstract

In modern age, the increasing complexity of computation and communication technology is leading us towards the necessity of new paradigm. As a result, unconventional approach like DNA coding theory is gaining considerable attention. The storage capacity, information processing and transmission properties of DNA molecules stimulate the notion of DNA coding theory as well as DNA cryptography. In this paper we generate DNA codeword using DNA linear block codes which ensures the secure transmission of information. In the proposed code design strategy DNA-based XOR operation (DNAX) is applied for effective construction of DNA codewords which are quadruples generated over the set of alphabets consisting of four DNA bases adenine, thymine, guanine, and cytosine. By worked out examples we explain the use of generator matrix and parity check matrix in encryption and decryption of coded data in the form of short single stranded DNA sequences. The newly developed technique can detect as well as correcting error in transmission of DNA codewords through biological channels from sender to the intended receiver. Through DNA coding theory we are expanding the paths towards data compression and error correction in the form of DNA strands. This leads us towards a broader domain of DNA cryptography.

****Corresponding Author***: *Mandrita Mondal, Indian Statistical Institute, 203, B.T. Road, Kolkata 700108, India; E-mail: mandritamondal@gmail.com*

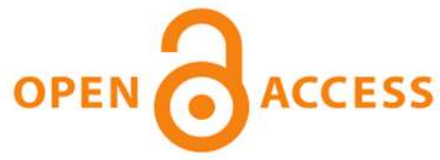

## 1. Introduction

In the modern era of e-business and e-commerce the protection of confidentiality, integrity, and availability (CIA triad) of stored information as well as of transmitted data is very crucial. Cryptography is the keystone of the data security system. The ancient techniques of cryptography have been evaluated by passing years by applying mathematics and logic to design strong encryption methodologies. The wide world of cryptography can be described as coding theory which includes data compression, error-correcting codes, and cryptography. Coding theory is the core of computation and communication. In modern age the increasing complexity of technology is leading us towards the necessity of new paradigm. As a result, unconventional approaches to coding theory have been developing from recent past and DNA coding theory is gaining considerable attention.

DNA molecules, which can be defined as the code of life, are being used in different technological and computing aspects apart from their biological functions; for example, use of DNA microarray in disease diagnostics; use of recombinant DNA technology in gene therapy and production of therapeutic proteins; use of synthetic DNA strands in DNA computing and nanotechnology viz. development of structural and dynamic autonomous DNA devices through programmed hybridization of complementary DNA sequences, solving challenging combinatorial problems and predicting consequence of logical reasoning and decision making problems by the manipulation of DNA strand by standard operations. Thus, in past several years a gradual and steady paradigm shift is occurring from silicon to carbon. This has been initiated years ago when Richard P. Feynman delivered the seminal lecture, “There’s Plenty of Room at the Bottom”, at the annual meeting of American Physical Society at Caltech in 1959 [1]. During his lecture, he mentioned about handling nanoparticles like DNA molecules and quantum molecules for computation. In 1987, Tom Head first merged molecular biology with formal language theory [2]. Finally, in 1994 Leonard Adleman explored the possibility of computation directly with molecules by solving seven-point Hamiltonian Path Problem [3] by DNA computing.

The storage capacity, information processing and transmission properties of DNA molecules inspire the idea of DNA coding theory as well as DNA cryptography. It is a rapid emerging unconventional methodology which combines the chemical characteristics of biological DNA sequences with classical techniques to ensure non-vulnerable transmission of data. The advantages of DNA computing over conventional silicon-based computation are:

1. **Massive parallelism:** In wet lab environment parallelly working $10^{18}$ processors can be handled by DNA computer.
2. **Potential for information storage:** DNA molecules require a trillion times less space to store same amount of data as existing storage media.
3. **Speed:** Though the elementary operations of DNA are slower, but because of the enormous parallelism, it can acquire more than 100,000 times of the speed of fastest supercomputer existing today.
4. **Energy efficiency:** It requires $10^{9}$ times less energy than conventional computer.

The technology applied in DNA cryptography for transmission of data are not encoded using mathematical operations, thus, it cannot be hacked easily. In this paper we design short single stranded DNA sequences, termed as DNA codewords, which can store, transmitting and retrieving secret information. Single stranded DNA sequences (i.e., oligonucleotides) are consist of quaternary sequences having four DNA bases i.e. A (adenine), T (thymine), C (cytosine) and G (guanine). Thus, the DNA codewords of a fixed length are generated over the set of four alphabets, $\Sigma_{DNA} = \{A, T, G, C\}$.

Different research works are being performed across the globe either to enhance the available methodologies or to propose innovative and novel approaches supporting the inevitable paradigm shift, from silicon to carbon. As the traditional coding theory is the pillar of modern information and communication technology, thus, it can be stated that DNA coding theory will be the base of DNA cryptography as well as DNA computation in near future. Designing set of DNA codewords for coding algorithms of DNA computation was proposed in [4-6]. Marathe et al. constructed combinatorial DNA codeword [7]. In [8] Milenkovic and Kashyap generated DNA words by which complex secondary structure formation can be avoided. Research works have also been conducted proposing the approaches of fabricating DNA codeword which can overcome the combinatorial constraints (GC-content constraint, reverse complement constraint, reverse constraint, Hamming distance constraint) [9-12], thermodynamic constraints (melting temperature constraint, free energy constraint, energy minimization constraint) [8,13-15] and application oriented constraints (run length constraint, correlated-uncorrelated constraint) [16]. The codes exploring the error-correcting properties of DNA molecules are developed in [17-19]. The use of block codes and convolution codes in prokaryotic translation initialization process was proposed by May et al. [20]. In [21] Rosen illustrated methodology for detection of error-correction coding structure in the nucleotide sequence. Battail [22] demonstrated the model of genomic error-correcting codes and proposed nested codes which is a layered structure for protection of information. In early course of research Forsdyke compared the non-coding intervening sequences, i.e., introns, with the noise affecting the transmission of signal in electronic medium [23]. He explained the possible role of introns in correcting errors of coding nucleotide sequence.

In the proposed work we generate DNA codewords using DNA *(n,k)* linear block codes in informational way which ensures the secure transmission of information. It is the fusion of classical coding theory and DNA computing technology. In the proposed code design strategy DNA-based XOR operation (DNAX) is applied for effective construction of DNA codewords. By worked out examples we explain the use of generator matrix and parity check matrix in encryption and decryption of coded data in the form of short single stranded DNA sequences. The newly developed technique can detect as well as correcting error in transmission of DNA codewords from sender to the intended receiver.

## 2. Preliminary Concepts

Before delving into deep of the proposed methodology, in this section we will discuss the basic concepts of linear block codes and DNA coding theory.

### 2.1. Linear block code

In coding theory [24], a set of fixed length words having well-defined mathematical property is termed as block code. Each word in the block code can be defined as codeword. A codeword consists of information bits and parity bits. Information bits carry the actual information and in contrary parity bits carry no information but ensure the proper structure required by the block code. The encoder maps the block of information bits i.e., information word into block of codeword. If n-bits of information words are coded into $k$-bits of codewords, it is termed as *(n,k)* block code, where $n > k$. Apart from the information bits, the extra bits in the codeword are termed as parity bits (*r*) which are determined by the encoder. It can be represented by $r = n - k$. Parity bits are placed arbitrarily in the codeword. In systematic code the information bits are kept together in the codeword so that they can be readily identified. Otherwise, the code is termed as non-systematic code. The schematic diagram of block encoder and codewords in *(n,k)* systematic block code is presented in Figure 1. In block codes the encoder is memoryless which means the output depends only on current *k*-bits of data block, not on the previous blocks. For binary *(n,k)* block code the set of codewords contains $2^k$ codewords of length $n$.

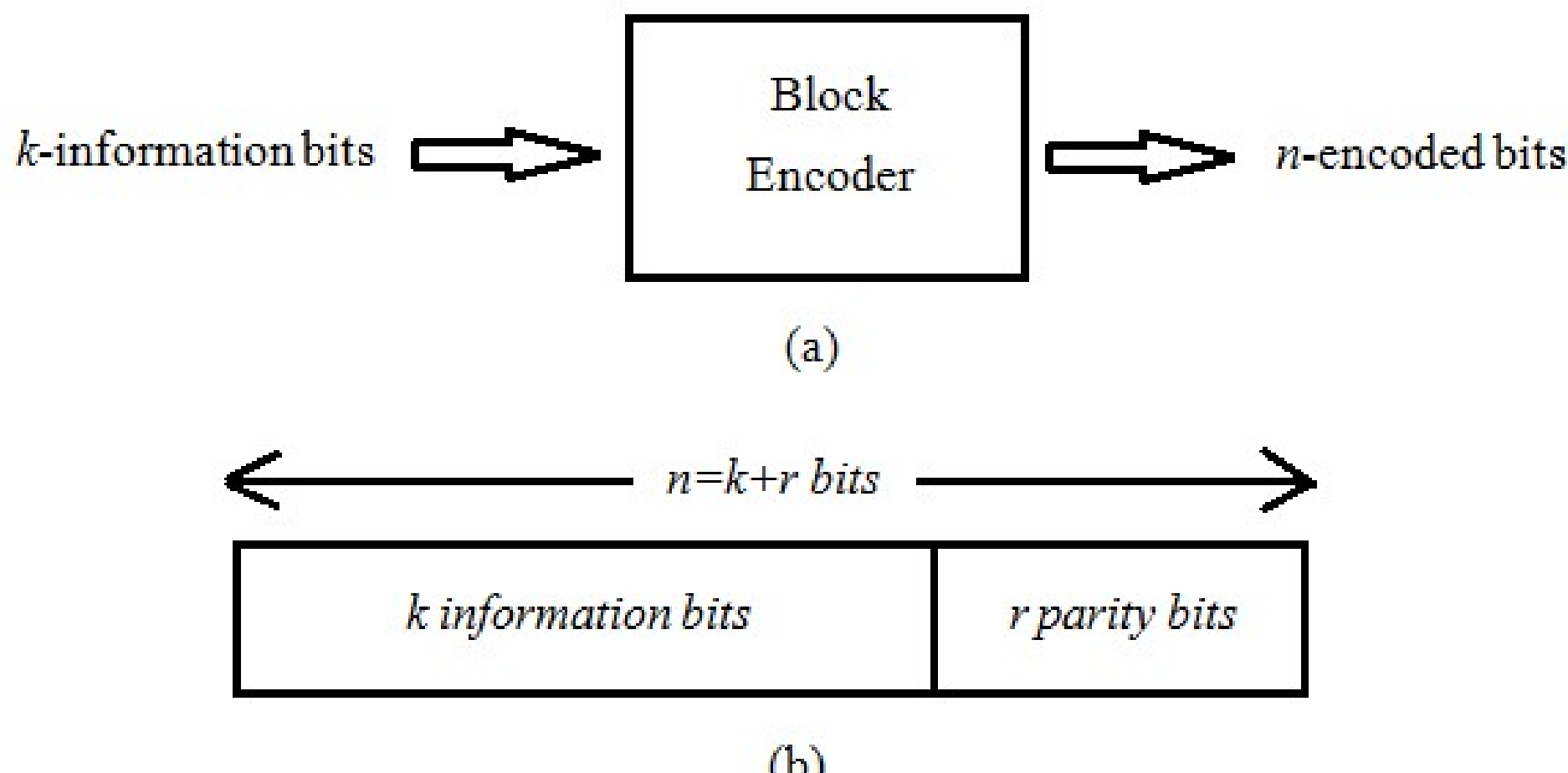

**Figure 1:** *(a) block encoder (b) codeword of (n,k) systematic block code.*

A block code can be established as *(n,k)* linear code if the linear combination of any two codewords from the set is also a codeword. Let, $c_{(n,k)} = \{c_1, c_2, \ldots\ldots, c_n\}$ is a set of n codewords and $c_x, c_y, c_z \in c$. Then, c is a linear block code if,

$$c_z = c_x \oplus c_y \qquad [1]$$

The occurrence of errors on a communication channel while transmitting data can be detected and corrected by linear block code. It is basically an error-correcting code.

**Example 1:** The information words and the corresponding codewords of (7,4) linear block code, denoted as $c_{(7,4)}$, are shown in Table 1. Let, the 4-bit information word is represented by $u = u_0\, u_1 u_2 u_3$ and the corresponding 7-bit codeword is represented by $c = v_0\, v_1 v_2 v_3 v_4 v_5 v_6$, where $c \in c_{(7,4)}$. The codewords of $c_{(7,4)}$ presented in Table 1 are generated by the following set of equations:

$$\left.\begin{aligned} v_0 &= u_0 \oplus u_2 \oplus u_3 \\ v_1 &= u_0 \oplus u_1 \oplus u_2 \\ v_2 &= u_1 \oplus u_2 \oplus u_3 \\ v_3 &= u_0 \\ v_4 &= u_1 \\ v_5 &= u_2 \\ v_6 &= u_3 \end{aligned}\right\} \qquad [2]$$

**Table 1:** *(7,4) linear block code.*

| Information word (k = 4) | Codeword (n = 7) |
|---|---|
| (0000) | (0000000) |
| (1000) | (1101000) |
| (0100) | (0110100) |
| (1100) | (1011100) |
| (0010) | (1110010) |
| (1010) | (0011010) |
| (0110) | (1000110) |
| (1110) | (0101110) |
| (0001) | (1010001) |
| (1001) | (0111001) |
| (0101) | (1100101) |
| (1101) | (0001101) |
| (0011) | (0100011) |
| (1011) | (1001011) |
| (0111) | (0010111) |
| (1111) | (1111111) |

The block code discussed in Example 1 is linear if the linear combination i.e., the modulo-2 sum of two arbitrarily chosen codewords from Table 1 is also a codeword (Equation 1).

**Example 2:** Two arbitrarily chosen codewords from $c_{(7,4)}$ are $c_4 = (1011100)$ and $c_{11} = (1100101)$. The linear combination of $c_4$ and $c_{11}$ is:

$$c_4 \oplus c_{11} = (1011100) \oplus (1100101) = (0111001) = c_{10}$$

As the linear combination of $c_4$ and $c_{11}$ is also a codeword i.e., $c_{10}$, then it can be concluded that the block code illustrated in Example 1 is linear.

### 2.1.1. Generator matrix

In coding theory, linear code is said to be the row space of its generator matrix which can be shown by the following equation.

$$c = uG \quad [3]$$

where, $c$ is a codeword.
$u$ is information word.
$G$ is the generator matrix for *(n,k)* linear block code having the size $k \times n$. The generator matrix can be expressed as:

$$G = \begin{bmatrix} \mathbf{g}_0 \\ \mathbf{g}_1 \\ \vdots \\ . \\ \mathbf{g}_{k-1} \end{bmatrix} = \begin{bmatrix} g_{0,0} & g_{0,1} & g_{0,2} & \cdots & g_{0,n-1} \\ g_{1,0} & g_{1,1} & g_{1,2} & \cdots & g_{1,n-1} \\ \cdot & \cdot & \cdot & \cdots & \cdot \\ \vdots & \vdots & \vdots & \vdots & \vdots \\ g_{k-1,0} & g_{k-1,1} & g_{k-1,2} & \cdots & g_{k-1,n-1} \end{bmatrix} \quad [4]$$

**Example 3:** The codewords of $c_{(7,4)}$ can be represented by the following equation;

$$c = [v_0\ v_1 v_2 v_3 v_4 v_5 v_6] = [u_0\ u_1 u_2 u_3] \begin{bmatrix} \mathbf{g}_0 \\ \mathbf{g}_1 \\ \mathbf{g}_2 \\ \mathbf{g}_3 \end{bmatrix}$$

$$= [u_0\ u_1 u_2 u_3] \begin{bmatrix} g_{0,0} & g_{0,1} & g_{0,2} & g_{0,3} & g_{0,4} & g_{0,5} & g_{0,6} \\ g_{1,0} & g_{1,1} & g_{1,2} & g_{1,3} & g_{1,4} & g_{1,5} & g_{1,6} \\ g_{2,0} & g_{2,1} & g_{2,2} & g_{2,3} & g_{2,4} & g_{2,5} & g_{2,6} \\ g_{3,0} & g_{3,1} & g_{3,2} & g_{3,3} & g_{3,4} & g_{3,5} & g_{3,6} \end{bmatrix} \quad [5]$$

The generator matrix of the linear block code $c_{(7,4)}$ is;

$$G = \begin{bmatrix} 1 & 1 & 0 & 1 & 0 & 0 & 0 \\ 0 & 1 & 1 & 0 & 1 & 0 & 0 \\ 1 & 1 & 1 & 0 & 0 & 1 & 0 \\ 1 & 0 & 1 & 0 & 0 & 0 & 1 \end{bmatrix} \quad [6]$$

Therefore, if an information word from the $c_{(7,4)}$ is (1001), then the corresponding codeword can be generated by Equation 3.

$$c = [1001] \begin{bmatrix} \mathbf{g}_0 \\ \mathbf{g}_1 \\ \mathbf{g}_2 \\ \mathbf{g}_3 \end{bmatrix} = [1.\mathbf{g}_0 + 0.\mathbf{g}_1 + 0.\mathbf{g}_2 + 1.\mathbf{g}_3] = [(1101000) + (1010001)] = (0111001) \quad [7]$$

### 2.1.2. Linear systematic block code

The systematic structure of the linear block code (Figure 1b) has two parts; information bit part (contains k unaltered information bits) and parity bit or redundant checking part (contains (*n-k*) parity-check bits). The four bits at the rightmost part of the codewords of $c_{(7,4)}$ (in Table 1) are identical to the information bits. The linear systematic form can be presented by generator matrix, part of which is identity matrix. The general systematic form of generator matrix of size $k \times n\text{-}k$ is shown in Equation 8. P is $k \times n\text{-}k$ parity check matrix and $I_k$ is the $k \times k$ identity matrix.

$$G = \begin{bmatrix} \mathbf{g}_0 \\ \mathbf{g}_1 \\ \vdots \\ . \\ \mathbf{g}_{k-1} \end{bmatrix} = \left[ \underbrace{\begin{matrix} p_{0,0} & p_{0,1} & \cdots & p_{0,n-k-1} \\ p_{1,0} & p_{1,1} & \cdots & p_{1,n-k-1} \\ \cdot & \cdot & \cdots & \cdot \\ \cdot & \cdot & \cdots & \cdot \\ p_{k-1,0} & p_{k-1,1} & \cdots & p_{k-1,n-k-1} \end{matrix}}_{(k \times n-k)P\ matrix} \middle| \underbrace{\begin{matrix} 1 & 0 & \cdots & 0 \\ 0 & 1 & \cdots & 0 \\ \cdot & \cdot & \cdots & \cdot \\ \cdot & \cdot & \cdots & \cdot \\ 0 & 0 & \cdots & 1 \end{matrix}}_{(k \times k)\ identity\ matrix} \right] = [P|I_k] \quad [8]$$

Here, $p_{i,j} = 0\ or\ 1\ for\ 0 \leq i < k\ and\ 0 \leq j < n-k$. Therefore, from Equation 5 and 8, the parity check equations can be written as Equation 9a and 9b.

$$v_{n-k+i} = u_i\ where, 0 \leq i < k \quad [9a]$$

$$v_j = u_0 p_{0,j} + u_1 p_{1,j} + \cdots + u_{k-1} p_{k-1,j} \; where, 0 \leq j < n - k \qquad [9b]$$

**Example 4:** The generator matrix of $c_{(7,4)}$ (Equation 6) can be written in systematic form (Equation 10) following the Equation 8.

$$G = \left[ \underbrace{\begin{matrix} 1 & 1 & 0 \\ 0 & 1 & 1 \\ 1 & 1 & 1 \\ 1 & 0 & 1 \end{matrix}}_{(4\times3) P\ matrix} \middle| \underbrace{\begin{matrix} 1 & 0 & 0 & 0 \\ 0 & 1 & 0 & 0 \\ 0 & 0 & 1 & 0 \\ 0 & 0 & 0 & 1 \end{matrix}}_{(4\times4)\ identity\ matrix} \right] = [P|I_4] \qquad [10]$$

Therefore, from Equation 5 and 10, Equation 11 can be extracted.

$$c = [v_0 \; v_1 v_2 v_3 v_4 v_5 v_6] = [u_0 \; u_1 u_2 u_3] \left[ \begin{array}{ccc|cccc} 1 & 1 & 0 & 1 & 0 & 0 & 0 \\ 0 & 1 & 1 & 0 & 1 & 0 & 0 \\ 1 & 1 & 1 & 0 & 0 & 1 & 0 \\ 1 & 0 & 1 & 0 & 0 & 0 & 1 \end{array} \right] \qquad [11]$$

The set of equations in Equation 2 are the parity check equations of $c_{(7,4)}$ which can be derived from Equations 9a and 9b.

### 2.1.3. Parity check matrix

In linear block code, the $k\times n$ generator matrix has $k$ linearly independent rows. Another significant matrix is *parity check matrix (H)* of size $n\text{-}k\times n$ with $n\text{-}k$ linearly independent rows. It can be said that any vector in the row space of $G$ is orthogonal to the rows of $H$. The criteria of the of an $n$-tuple word $c$, generated by the generator matrix $G$, being a codeword is given in Equation 12.

$$c.H^T = 0 \qquad [12]$$

From Equation 3 it can be written that,

$$uG.H^T = 0 \qquad [13]$$

Equation 13 implies that the rows of $G$ matrix and $H$ matrix are orthogonal to each other i.e., $H$ lies in the null space of $G$. The general systematic form of parity check matrix is given in Equation 14.

$$H = [I_{n-k}|P^T] = \left[ \begin{array}{cccc|cccc} 1 & 0 & \ldots & 0 & p_{0,0} & p_{1,0} & \ldots. & p_{k-1,0} \\ 0 & 1 & \ldots & 0 & p_{0,1} & p_{1,1} & \ldots. & p_{k-1,1} \\ . & . & \ldots & . & . & . & \ldots. & . \\ . & . & \ldots & . & . & . & \ldots. & . \\ 0 & 0 & \ldots & 1 & p_{0,n-k-1} & p_{1,n-k-1} & \ldots. & p_{k-1,n-k-1} \end{array} \right] \qquad [14]$$

**Example 5:** The parity check matrix of $c_{(7,4)}$ is given below;

$$H = [I_3|P^T] = \left[ \begin{array}{ccc|cccc} 1 & 0 & 0 & 1 & 0 & 1 & 1 \\ 0 & 1 & 0 & 1 & 1 & 1 & 0 \\ 0 & 0 & 1 & 0 & 1 & 1 & 1 \end{array} \right] \qquad [15]$$

Therefore, from Equations 5, 12 and 15 the criteria of a word being a codeword of $c_{(7,4)}$ is expressed as Equation 16.

$$[v_0\ v_1 v_2 v_3 v_4 v_5 v_6]\begin{bmatrix}1 & 0 & 0\\0 & 1 & 0\\0 & 0 & 1\\1 & 1 & 0\\0 & 1 & 1\\1 & 1 & 1\\1 & 0 & 1\end{bmatrix} = [0 \quad 0 \quad 0] \quad [16]$$

### 2.1.4. Minimum distance and error-correcting capability of linear block code

The minimum distance, $d_{min}$, of a linear block code can be defined as,

$$d_{min} = \min\{d(c_x, c_y): c_x, c_y \in c, c_x \neq c_y\} \quad [17]$$

As the linear combination of two codes is also a codeword (Equation 1), Equation 18 can be drawn.

$$d_{min} = \min\{d(c_x, c_y): c_x, c_y \in c, c_x \neq c_y\} = \min\{w(c_z): c_z \in c, c_z \neq 0\} = w_{min} \quad [18]$$

where, $w(c_z)$ is Hamming weight of $c_z$.
$w_{min}$ is the minimum weight of the linear block code.

Thus, it can be concluded that, the minimum distance of a linear block code is equal to the minimum weight of its non-zero codewords. For example, in $c_{(7,4)}$ $d_{min} = w_{min} = 3$.

The *random-error-correcting capability* ($t$) of a linear block code is defined by the following expression,

$$t = [(d_{min} - 1)/2] \quad [19]$$

Linear block code can correct all the error patterns of $t$ or fewer errors. For $c_{(7,4)}$, $t$= 1.

### 2.1.5. Syndrome and error detection

Let, $c = (v_0, v_1, \dots\dots, v_{n-1})$ be a codeword from binary *(n,k)* linear block code with generator matrix *G* and parity check matrix *H*. The codeword *c* is transmitted through binary symmetric channel and the intended receiver receives $r = (r_0, r_1, \dots\dots, r_{n-1})$ as the output of the transmission of data. If the transmission channel is noisy *r* can be different from *c*. The relation between the transmitted codeword and the codeword at the receiving end can be illustrated by the following expression.

$$\begin{aligned} r &= (r_0, r_1, \dots\dots, r_{n-1}) = c + e = (v_0, v_1, \dots\dots, v_{n-1}) + (e_0, e_1, \dots\dots, e_{n-1}) \\ &= (v_0 + e_0, v_1 + e_1, \dots\dots v_{n-1} + e_{n-1}) \end{aligned} \quad [20]$$

where, $e = (e_0, e_1, \dots\dots, e_{n-1})$ is the binary *error pattern* or *error vector*. Here, modulo-2 sum has been considered.

The occurrence of error must be detected so that the decoder can take relevant action. The *n*-tuple $e_i$ can be expressed as,

$$e_i = \begin{cases} 1 \; if \; r_i \neq v_i \\ 0 \; if \; r_i = v_i \end{cases} \text{ where, } 0 < i \leq n-1 \quad [21]$$

$e_i = 1$ indicates that the $i^{th}$ position of $r$ has an error. Error detection can be achieved by computing *(n-k)*-tuple $S$, termed as *syndrome* which can be expressed by Equation 22

$$S = (S, S_1, \dots \dots, S_{n-k-1}) = rH^T \quad [22]$$

where, the size of $r$ is $n$×$n$ and $H$ matrix is $n$×$n$-$k$.

If $S \neq 0$, then $r$ is not a codeword and transmission error has been detected. If $S = 0$, no error has been detected. If $r$ is a codeword other than the transmitted codeword, then an *undetected error* occurs. This happens when the error pattern is also a codeword and following Equation 1 $e$ transforms one codeword in some other codeword. Thus, it can be concluded that, $S$ depends only on $e$, not on $r$.

$$S = rH^T = (c + e).H^T = c.H^T + e.H^T = e.H^T \text{ (by Equation 12)} \quad [23]$$

Therefore,

$$S_j = e_j + e_{n-k}p_{0,j} + e_{n-k+1}p_{1,j} + \cdots + e_{n-1}p_{k-1,j} \quad [24]$$

where, $0 \leq j < n - k$.

From the set of *(n-k)* syndrome equations (Equation 24) there are $2^n$ possible solutions among which only one solution represents the true error pattern. To minimize the probability of a decoding error, the most probable error pattern that satisfies the above equations is chosen as the true error vector. The error vector is used to correct the specific number of errors in transmitted codeword.

After explaining the general concepts of linear block code, in the next subsection we will discuss a brief overview on DNA coding theory.

## 2.2. DNA coding theory

DNA molecules have the distinctive property to store, process and transmit data which stimulates the notion of DNA coding theory as well as DNA cryptography. New paradigm for non-vulnerable coding and fast computing is being evolved through the amalgamation of biological science and computational science. DNA codewords, generally short single stranded DNA sequences i.e., oligonucleotides, can store, transmitting and retrieving information. We consider the DNA codewords of fixed length which are quadruples and generated over the set of four alphabets, $\Sigma_{DNA} = \{A, T, G, C\}$. *A* represents *adenine*, *T* represents *thymine*, *C* represents *cytosine* and *G* represents *guanine*.

One of the unique properties of DNA molecule is *complementary base-pairing* or *Watson-Crick pairing*. Adenine is the complementary base to thymine and guanine is the complementary base to cytosine which can be symbolically represented as,

$$A^C = T;\ T^C = A;\ G^C = C;\ C^C = G \quad [25]$$

The non-covalent *H*-bonds formed between the complementary bases holds two single DNA strands together in *antiparallel* orientation. The base pairing of two strands, with opposite polarity, in double stranded DNA sequences holds them together i.e., the base at 5' end of one strand is paired with the base at 3' end of the other strand. For example, the complementary strand of DNA sequence 5'-AGATCTA-3' is 3'-TCTAGAT-5'.

Now we adopt the following convention: the complementary base of the single base $\sigma_i$ of a DNA sequence is $\bar{\sigma}_i$, where $0 < i \leq n-1$ for a DNA strand with length $n$. Let, $\sigma = \sigma_0\ \sigma_1\sigma_2 \ldots\ldots \sigma_{n-1}$ (5′ to 3′ direction) is *n* bases long DNA sequence. The *reverse sequence* of $\sigma$ can be represented as, $\sigma^R = \sigma_{n-1}\ \sigma_{n-2} \ldots\ldots \sigma_1\sigma_0$ (5′ to 3′ direction) and the *complementary sequence* of $\sigma$ is, $\sigma^C = \bar{\sigma}_0\ \bar{\sigma}_1\bar{\sigma}_2 \ldots\ldots \bar{\sigma}_{n-1}$ (5′ to 3′ direction). Single stranded DNA can hybridize to its reverse complementary sequence and form double stranded DNA sequence. Therefore, the reverse complementary sequence of $\sigma$ can be denoted as, $\sigma^{RC} = \bar{\sigma}_{n-1}\ \bar{\sigma}_{n-2} \ldots\ldots \bar{\sigma}_1\bar{\sigma}_0$ (5′ to 3′ direction). The double stranded DNA sequence is symbolized as $\begin{bmatrix} \sigma_0 & \sigma_1 & \ldots. & \sigma_{n-1} \\ \bar{\sigma}_0 & \bar{\sigma}_1 & \ldots. & \bar{\sigma}_{n-1} \end{bmatrix}$ in which the first strand is presented in 5′ to 3′ direction and the reverse complementary strand is presented in 3′ to 5′ direction.

Various approaches have been proposed by the researchers for generation of DNA codewords having finite length, specific distance and satisfying combinatorial, thermodynamic, and application-based constraints. In the next section we will discuss our proposed design strategy for construction of *DNA (n,k) block codes* which is error-detecting and error-correcting coding methodology.

## 3. Designing DNA Codewords by Linear Block Codes

Before transmission of data, *cipher text* is generated from *plain text* to protect it from the third parties i.e., adversaries. In this paper we focus on the construction of *DNA (n,k) block codes*. To achieve this, first plain text must be converted into DNA sequence by following one of the two encoding methodologies.

1. **Conversion of Binary into DNA sequence:** Plaintext, which is supposed to be transmitted, is converted into binary sequence from ASCII values. DNA sequence can be encoded from the converted binary form by applying one of these proposals [25-27]. A simple example of encoding plan can be explained as; *00* is encoded as *A*, *01* is encoded as *T*, *10* is encoded as *C* and *11* is encoded as *G*.
2. **Conversion of Plaintext by Encoding Manual:** The alphabets of the plaintext are converted into fixed length DNA oligonucleotides following a predefined manual containing encoding table [28,29].

### 3.1. DNA *(n,k)* linear block code

In DNA block code, first the plain text is converted into binary form and again the binary sequence is used to generate DNA string as described earlier. The generated DNA sequence is treated as the information sequence which is partitioned into message blocks of *k*-information bases each represented by,

$$u^{DNA} = (u_0^{DNA}\ u_1^{DNA}u_2^{DNA} \ldots\ldots u_{k-1}^{DNA}) \qquad [26]$$

**Definition 1:** Following the classical coding theory as described in section 2.1, *DNA (n,k) block codes* can be defined as mapping of *k*-bases long DNA block into *n*-bases long DNA codeword. The DNA codeword $c^{DNA}$ can be represented as,

$$c^{DNA} = (v_0^{DNA}\ v_1^{DNA} v_2^{DNA} \dots.. v_{n-1}^{DNA}) \quad [27]$$

**Definition 2:** The extra $r$ bases, where $r = n - k$, in the DNA codeword are termed as *parity bases*.

As DNA sequences are quadruples generated over the set of alphabets i.e., $\Sigma_{DNA} = \{A, T, G, C\}$, DNA *(n,k)* block code contains a set of $4^k$ codewords. Following Example 1, we illustrate DNA (7,4) block code in Example 6.

**Example 6:** In DNA (7,4) block code 4-bases long information DNA block is mapped into 7-bases long DNA codeword. From Equation 26, the information block can be presented as, $u^{DNA} = u_0^{DNA}\ u_1^{DNA} u_2^{DNA} u_3^{DNA}$ and the corresponding 7-base codeword can be written as, $c^{DNA} = v_0^{DNA}\ v_1^{DNA} v_2^{DNA} v_3^{DNA} v_4^{DNA} v_5^{DNA} v_6^{DNA}$where,$c^{DNA} \in c_{(7,4)}^{DNA}$. Few arbitrarily chosen DNA codewords from the set of 256 ($4^4$) codewords in $c_{(7,4)}^{DNA}$ are shown in Table 2. Following Equation 2, the codewords are generated by the set of equations (Equation 28) given below.

$$\left.\begin{aligned} v_0^{DNA} &= u_0^{DNA} \oplus u_2^{DNA} \oplus u_3^{DNA} \\ v_1^{DNA} &= u_0^{DNA} \oplus u_1^{DNA} \oplus u_2^{DNA} \\ v_2^{DNA} &= u_1^{DNA} \oplus u_2^{DNA} \oplus u_3^{DNA} \\ v_3^{DNA} &= u_0^{DNA} \\ v_4^{DNA} &= u_1^{DNA} \\ v_5^{DNA} &= u_2^{DNA} \\ v_6^{DNA} &= u_3^{DNA} \end{aligned}\right\} \quad [28]$$

**Table 2:** *Examples of DNA (7,4) block codes.*

| Information bases ($k$ = 4) | DNA Codeword ($n$ = 7) |
|---|---|
| (*ATCA*) | (*CGGATCA*) |
| (GCTG) | (*TAAGCTG*) |
| (TGGC) | (*ATCTGGC*) |
| (CATC) | (*TGGCATC*) |
| (*TCAG*) | (*CGTTCAG*) |

**Definition 3:** A DNA block code can be established as *DNA (n,k) linear code* if the linear combination of any two codewords from the set is also a codeword. Let, $c^{DNA} = \{c_1^{DNA}, c_2^{DNA}, \dots.., c_n^{DNA}\}$ is a set of $n$ DNA codewords and $c_x^{DNA}, c_y^{DNA}, c_z^{DNA} \in c^{DNA}$. Then, $c^{DNA}$ is a linear block code if,

$$c_z^{DNA} = c_x^{DNA} \oplus c_y^{DNA} \quad [29]$$

The linear combination of two DNA codewords indicates the *DNA-based XOR operation (DNAX)* [30]. Table 3 shows the DNAX operation which mimics binary XOR operation and has uniqueness and reflexive properties.

**Table 3:** *DNA-based XOR operation (DNAX).*

| DNAX | A | G | C | T |
|---|---|---|---|---|
| **A** | *A* | *G* | *C* | *T* |
| **G** | *G* | *A* | *T* | *C* |
| **C** | *C* | *T* | *A* | *G* |
| **T** | *T* | *C* | *G* | *A* |

The DNA block code discussed in Example 6 is linear if the linear combination i.e., the DNAX of two arbitrarily chosen DNA codewords from Table 2 is also a codeword (Equation 29).

**Example 7:** Two arbitrarily chosen DNA codewords from $c_{(7,4)}^{DNA}$ are (*CGGATCA*) and (*ATCTGGC*). Performing DNAX of these two codewords,

(*CGGATCA*) ⊕ (*ATCTGGC*) = (*CCTTCTC*)

The resultant sequence (*CCTTCTC*) is also a codeword, the corresponding information sequence of which is (*TCTC*). This can be verified using Equation 28. Thus, we can conclude that the DNA block code illustrated in Example 6 is linear.

### 3.1.1. Use of generator matrix for construction of DNA codewords

To construct DNA codewords by generator matrix, we need to define two mathematical operations between DNA bases and binary bits: (1) *multiplication* and (2) *DNAXOR*.

**Definition 4:** Multiplication between DNA base, denoted by *x*, and binary bit (0 or 1) is defined by the following set of equations.

$$\left.\begin{matrix} x.1 = x \\ x.0 = 0 \end{matrix}\right\} \quad [30]$$

**Definition 5:** DNAXOR between DNA base, denoted by *x*, and binary bit (0 or 1) is defined by the following set of equations.

$$\left.\begin{matrix} x \oplus 1 = 0 \\ x \oplus 0 = x \end{matrix}\right\} \quad [31]$$

Like classical coding theory, DNA *(n,k)* linear block code can also be defined by *k* × *n* generator matrix. Following Equation 3, DNA linear block code is also said to be the row space of its generator matrix. The elements of DNA generator matrix are binary bits and DNA codewords are constructed by multiplication and DNAXOR operation defined in Definitions 4 and 5. The following expression represents the construction of DNA codewords from generator matrix.

$$c^{DNA} = u^{DNA}G \quad [32]$$

The generator matrix can be constructed from Equations 30 and 31 and set of equations those are used to generate the codeword from corresponding information word (for example, Equation 28).

**Example 8:** $c_{(7,4)}^{DNA}$ is illustrated in Example 6. From Equations 30 and 31 the corresponding generator matrix can be constructed. As we know that,

$$[v_0^{DNA}\ v_1^{DNA} v_2^{DNA} v_3^{DNA} v_4^{DNA} v_5^{DNA} v_6^{DNA}] = [u_0^{DNA}\ u_1^{DNA} u_2^{DNA} u_3^{DNA}] \begin{bmatrix} g_{0,0} & g_{0,1} & g_{0,2} & g_{0,3} & g_{0,4} & g_{0,5} & g_{0,6} \\ g_{1,0} & g_{1,1} & g_{1,2} & g_{1,3} & g_{1,4} & g_{1,5} & g_{1,6} \\ g_{2,0} & g_{2,1} & g_{2,2} & g_{2,3} & g_{2,4} & g_{2,5} & g_{2,6} \\ g_{3,0} & g_{3,1} & g_{3,2} & g_{3,3} & g_{3,4} & g_{3,5} & g_{3,6} \end{bmatrix} \quad [33]$$

If $u^{DNA}$ = (*GCTG*) and the corresponding $c^{DNA}$ = (*TAAGCTG*), then Equation 34 can be written as,

$$[TAAGCTG] = [GCTG] \begin{bmatrix} g_{0,0} & g_{0,1} & g_{0,2} & g_{0,3} & g_{0,4} & g_{0,5} & g_{0,6} \\ g_{1,0} & g_{1,1} & g_{1,2} & g_{1,3} & g_{1,4} & g_{1,5} & g_{1,6} \\ g_{2,0} & g_{2,1} & g_{2,2} & g_{2,3} & g_{2,4} & g_{2,5} & g_{2,6} \\ g_{3,0} & g_{3,1} & g_{3,2} & g_{3,3} & g_{3,4} & g_{3,5} & g_{3,6} \end{bmatrix} \quad [34]$$

The constructed generator matrix for $c_{(7,4)}^{DNA}$ is,

$$G = \begin{bmatrix} 1 & 1 & 0 & 1 & 0 & 0 & 0 \\ 0 & 1 & 1 & 0 & 1 & 0 & 0 \\ 1 & 1 & 1 & 0 & 0 & 1 & 0 \\ 1 & 0 & 1 & 0 & 0 & 0 & 1 \end{bmatrix} \quad [35]$$

The generator matrix is in systematic form. In Example 9 we show that, if generation matrix and information sequence are given, then the corresponding DNA codeword can be constructed following Equation 32.

**Example 9:** If an information sequence from $c_{(7,4)}^{DNA}$ is (*TCAG*), then the corresponding codeword from generator matrix is deduced below.

$$\begin{aligned} c &= [\mathrm{TCAG}] \begin{bmatrix} \mathfrak{g}_0 \\ \mathfrak{g}_1 \\ \mathfrak{g}_2 \\ \mathfrak{g}_3 \end{bmatrix} = [T.\mathfrak{g}_0 \oplus C.\mathfrak{g}_1 \oplus A.\mathfrak{g}_2 \oplus G.\mathfrak{g}_3] \\ &= [T.(1101000) \oplus C.(0110100) \oplus A.(1110010) \oplus G.(1010001)] \\ &= [(TT0T000) \oplus (0CC0C00) \oplus (AAA00A0) \oplus (G0G000G)] \text{ (from Equation 30)} \\ &= [(TGCTC00) \oplus (AAA00A0) \oplus (G0G000G)] \text{ (from Equation 31 and Table 3)} \\ &= [(TGCTCA0) \oplus (G0G000G)] \\ &= [CGTTCAG] \end{aligned} \quad [36]$$

The deduced sequence (*CGTTCAG*) is the corresponding DNA codeword of DNA information sequence (*TCAG*) which can be verified from the last row of Table 2 in Example 6.

### 3.1.2. DNA parity check matrix

*DNA parity check matrix* ($H^{DNA}$), having the size $n\text{-}k \times n$ with $n\text{-}k$ linearly independent rows, is somehow similar to classical parity check matrix. Any vector in the in the row space of $G$ is orthogonal to the rows of $H^{DNA}$. The criteria of a quadruple being a DNA codeword is defined in Definition 6.

**Definition 6.** The quadruple $c^{DNA} = (v_0^{DNA}\ v_1^{DNA} v_2^{DNA} \ldots\ldots v_{n-1}^{DNA})$ is a DNA codeword if and only if it satisfies Equation 37.

$$c^{DNA}.(H^{DNA})^T = (A\ A \ldots.A) \quad [37]$$

From Equation 32 it can be written that,

$$u^{DNA}G.(H^{DNA})^T = (A\ A \ldots.A) \quad [38]$$

Therefore, the general systematic form of DNA parity check matrix is given in Equation 39.

$$H^{DNA} = [I_{n-k}^{DNA}|(P^{DNA})^T] = \begin{bmatrix} 1 & 0 & \ldots & 0 & p_{0,0}^{DNA} & p_{1,0}^{DNA} & \ldots. & p_{k-1,0}^{DNA} \\ 0 & 1 & \vdots & 0 & p_{0,1}^{DNA} & p_{1,1}^{DNA} & \ldots. & p_{k-1,1}^{DNA} \\ . & . & \ldots & . & . & . & \ldots. & . \\ . & . & \ldots & . & . & . & \ldots. & . \\ 0 & 0 & \ldots & 1 & p_{0,n-k-1}^{DNA} & p_{1,n-k-1}^{DNA} & \ldots. & p_{k-1,n-k-1}^{DNA} \end{bmatrix} \quad [39]$$

**Example 10:** The generator matrix of $c_{(7,4)}^{DNA}$ has been discussed in Example 8. Therefore, the corresponding $H^{DNA}$ is shown in Equation 40.

$$H^{DNA} = [I_3^{DNA}|(P^{DNA})^T] = \left[\begin{array}{ccc|cccc} 1 & 0 & 0 & 1 & 0 & 1 & 1 \\ 0 & 1 & 0 & 1 & 1 & 1 & 0 \\ 0 & 0 & 1 & 0 & 1 & 1 & 1 \end{array}\right] \quad [40]$$

The criteria of a quadruple being a codeword in $c_{(7,4)}^{DNA}$ is expressed as Equation 41.

$$[v_0\ v_1 v_2 v_3 v_4 v_5 v_6] \begin{bmatrix} 1 & 0 & 0 \\ 0 & 1 & 0 \\ 0 & 0 & 1 \\ 1 & 1 & 0 \\ 0 & 1 & 1 \\ 1 & 1 & 1 \\ 1 & 0 & 1 \end{bmatrix} = [A \quad A \quad A] \quad [41]$$

In Example 11 we prove that the criteria claimed in Definition 6 is true.

**Example 11:** We have arbitrarily chosen an information sequence of $c_{(7,4)}^{DNA}$ which is (*CATC*). The corresponding DNA codeword given in Table 2 is (*TGGCATC*). The sequence (*TGGCATC*) is DNA codeword if it satisfies Equation 41.

$$[\mathrm{TGGCATC}] \begin{bmatrix} 1 & 0 & 0 \\ 0 & 1 & 0 \\ 0 & 0 & 1 \\ 1 & 1 & 0 \\ 0 & 1 & 1 \\ 1 & 1 & 1 \\ 1 & 0 & 1 \end{bmatrix} = [(T \oplus C \oplus T \oplus C) \quad (G \oplus C \oplus A \oplus T) \quad (G \oplus A \oplus T \oplus C)] \\ = [A \quad A \quad A] \quad [42]$$

From Equation 42 it has been proved that (*TGGCATC*) is the codeword corresponding to the information word (*CATC*) in $c_{(7,4)}^{DNA}$.

**Example 12:** In Example 11 we have proved that the DNA sequence (*TGGCATC*) is codeword. In the present example we are considering a modified sequence in which a single base position of the above-mentioned DNA codeword has been altered. Let, the modified sequence is (TGG*T*ATC) in which *C* in fourth position has been replaced by *T*. Now applying Equation 37 we can verify if the altered DNA sequence is a codeword in $c_{(7,4)}^{DNA}$.

$$[\text{TGGTATC}]\begin{bmatrix}1 & 0 & 0\\0 & 1 & 0\\0 & 0 & 1\\1 & 1 & 0\\0 & 1 & 1\\1 & 1 & 1\\1 & 0 & 1\end{bmatrix} = [(T \oplus T \oplus T \oplus C) \quad (G \oplus T \oplus A \oplus T) \quad (G \oplus A \oplus T \oplus C)] \\ = [G \quad G \quad A] \qquad [43]$$

From Equation 43 it has been proved that (*TGGTATC*) is not a codeword in $c_{(7,4)}^{DNA}$.

So far, we have considered DNA (7,4) linear block code. In Example 13 we will explain the linear block coding using DNA (6, 3) linear block code i.e., $c_{(6,3)}^{DNA}$.

**Example 13:** Arbitrarily chosen DNA codewords from the set of 64 ($4^3$) codewords in $c_{(6,3)}^{DNA}$ are shown in Table 4. The set of equations (Equation 44), from which the DNA codewords are constructed, are given below.

$$\left.\begin{aligned} v_0^{DNA} &= u_1^{DNA} \oplus u_2^{DNA}\\ v_1^{DNA} &= u_0^{DNA} \oplus u_2^{DNA}\\ v_2^{DNA} &= u_0^{DNA} \oplus u_1^{DNA}\\ v_3^{DNA} &= u_0^{DNA}\\ v_4^{DNA} &= u_1^{DNA}\\ v_5^{DNA} &= u_2^{DNA} \end{aligned}\right\} \qquad [44]$$

**Table 4:** *Examples of DNA (6, 3) block codes.*

| Information bases ($k$ = 3) | DNA Codeword ($n$ = 6) |
|---|---|
| (*AAT*) | (*TTAAAT*) |
| (*AAC*) | (*CCAAAC*) |
| (*TTT*) | (*AAATTT*) |
| (*TTC*) | (*GGATTC*) |
| (*CGC*) | (*TATCGC*) |

Now we take the linear combination of two arbitrarily chosen DNA codewords from Table 4. Performing DNAX of these two codewords (*TTAAAT*) and (*GGATTC*);

(*TTAAAT*) ⊕ (*GGATTC*) = (*CCATTG*)

The resultant sequence (*CCATTG*) is also a codeword if it satisfies the criteria stated in Definition 6.

The generator matrix of $c_{(6,3)}^{DNA}$ is,

$$G = \begin{bmatrix} 0 & 1 & 1 & 1 & 0 & 0 \\ 1 & 0 & 1 & 0 & 1 & 0 \\ 1 & 1 & 0 & 0 & 0 & 1 \end{bmatrix}$$

The $H^{DNA}$ of $c^{DNA}_{(6,3)}$ is,

$$H^{DNA} = \begin{bmatrix} 1 & 0 & 0 & 0 & 1 & 1 \\ 0 & 1 & 0 & 1 & 0 & 1 \\ 0 & 0 & 1 & 1 & 1 & 0 \end{bmatrix}$$

Now,

$$[\text{CCATTG}].(H^{DNA})^T = [\text{CCATTG}] \begin{bmatrix} 1 & 0 & 0 \\ 0 & 1 & 0 \\ 0 & 0 & 1 \\ 0 & 1 & 1 \\ 1 & 0 & 1 \\ 1 & 1 & 0 \end{bmatrix} = [(C \oplus T \oplus G) \quad (C \oplus T \oplus G) \quad (A \oplus T \oplus T)] = [A \quad A \quad A] \qquad [45]$$

From Equation 45 it has been proven that the sequence (*CCATTG*) is a DNA codeword which is the linear combinations of two codewords (*TTAAAT*) and (*GGATTC*). Thus, $c^{DNA}_{(6,3)}$ is a linear block code.

### 3.1.3. Minimum distance and error-correcting capability of DNA *(n,k)* linear block code

The minimum distance of DNA linear block code ($d^{DNA}_{min}$) can be defined as,

$$d^{DNA}_{min} = \min\{d(c^{DNA}_x, c^{DNA}_y): c^{DNA}_x, c^{DNA}_y \in c^{DNA}, c^{DNA}_x \neq c^{DNA}_y\} \qquad [46]$$

For DNA linear block code, the minimum distance is equal to the minimum weight of its non-adenine (*non-A*) codewords. Following Equation 18, Equation 47 can be drawn that,

$$d^{DNA}_{min} = w^{DNA}_{min} \qquad [47]$$

where, $w^{DNA}_{min}$ is the minimum weight of the DNA linear block code which is the minimum count of the *non-A* bases of the corresponding codeword.

Thus, the *random-error-correcting capability* ($t^{DNA}$) of a DNA linear block code can be defined by,

$$t^{DNA} = [(d^{DNA}_{min} - 1)/2] \qquad [48]$$

### 3.1.4. Error syndromes in DNA linear block code

So far, we have studied the design strategy of DNA codewords by DNA *(n,k)* linear block codes. The codeword, which actually contains the data, is supposed to be securely transferred to the intended receiver. There are few existing methodologies by which DNA codeword can be transmitted from the sender to the recipient. We will discuss some techniques of transmission of encrypted DNA sequences.

Clelland et al. [31] proposed the transmission of coded sequences using DNA microdots. DNA microdots are microscopic DNA spots attached to a solid surface. Wong et al. [32] presented the idea of permanent storage of DNA codewords in the living host securely and allowing the organism to grow and multiply. This procedure ensures the protection of encrypted DNA sequences from the adverse circumstances, such as, fatal double strand break of DNA caused by extreme temperature and desiccation or rehydration; presence of DNA nucleases; ultraviolet ray, ionizing radiation; intentional attack by any individual etc. The preserved information can be recovered again.

Though these proposed methods of transmission of DNA codewords are supposed to be secure, but sometimes induced mutation can occur. It can alter certain base in the encrypted DNA sequence. Let $c^{DNA} = \left(v_0^{DNA}\ v_1^{DNA} v_2^{DNA} \ldots\ldots v_{n-1}^{DNA}\right)$ is a codeword generated in $c_{(n,k)}^{DNA}$ has been transmitted and the recipient has received the word $r^{DNA} = \left(r_0^{DNA}\ r_1^{DNA} r_2^{DNA} \ldots\ldots r_{n-1}^{DNA}\right)$. Now $r^{DNA}$ may differ from $c^{DNA}$ because of the induced mutation.

Equation 49 represents the relation between the transmitted codeword $c^{DNA}$ and the received codeword i.e., $r^{DNA}$.

$$\begin{aligned} r^{DNA} &= \left(r_0^{DNA}\ r_1^{DNA} \ldots\ldots r_{n-1}^{DNA}\right) = c^{DNA} \oplus e^{DNA} \\ &= \left(v_0^{DNA}\ v_1^{DNA} \ldots\ldots v_{n-1}^{DNA}\right) \oplus \left(e_0^{DNA}\ e_1^{DNA} \ldots\ldots e_{n-1}^{DNA}\right) \\ &= \left(v_0^{DNA} \oplus e_0^{DNA}, v_1^{DNA} \oplus e_1^{DNA}, \ldots\ldots v_{n-1}^{DNA} \oplus e_{n-1}^{DNA}\right) \end{aligned} \quad [49]$$

where, $e^{DNA}$ is the *DNA error pattern*.

**Definition 7:** *DNA syndrome*, $S^{DNA}$, is *(n-k)*-tuple by which transmission error in DNA linear block code can be detected. DNA syndrome can be expressed as,

$$S^{DNA} = (S_0^{DNA}, S_1^{DNA} \ldots\ldots S_{n-k-1}^{DNA}) = r^{DNA}(H^{DNA})^T \quad [50]$$

where, the size of $r^{DNA}$ is $1 \times n$ and the size of $H^{DNA}$ matrix is $n \times n\text{-}k$.

From Equation 37 it can be stated that if $S^{DNA} = r^{DNA}.(H^{DNA})^T = (A\ A \ldots. A)$, then no error has been detected and $r^{DNA}$ is a DNA codeword of DNA linear block code; otherwise it is not a codeword.

**Example 14:** Let $r^{DNA} = (GGATTC)$ in $c_{(6,3)}^{DNA}$. Then,

$$\begin{aligned} S^{DNA} &= [\mathrm{GGATTC}].(H^{DNA})^T = [\mathrm{GGATTC}]\begin{bmatrix} 1 & 0 & 0 \\ 0 & 1 & 0 \\ 0 & 0 & 1 \\ 0 & 1 & 1 \\ 1 & 0 & 1 \\ 1 & 1 & 0 \end{bmatrix} \\ &= [(G \oplus T \oplus C) \quad (G \oplus T \oplus C) \quad (A \oplus T \oplus T)] = [A \quad A \quad A] \end{aligned} \quad [51]$$

The Equation 51 proves that $r^{DNA} = (GGATTC)$ is a codeword in $c_{(6,3)}^{DNA}$.

**Example 15:** Assume that in transmission of DNA codewords (in $c^{DNA}_{(6,3)}$) an induced mutation occurs which leads to the alteration of a base in $c^{DNA} = (GGATTC)$. In second base position $G$ is replaced by $A$ and $r^{DNA}$ = (G**A**ATTC). Then,

$$S^{DNA} = [\text{GAATTC}].(H^{DNA})^T = [\text{GAATTC}]\begin{bmatrix}1 & 0 & 0\\ 0 & 1 & 0\\ 0 & 0 & 1\\ 0 & 1 & 1\\ 1 & 0 & 1\\ 1 & 1 & 0\end{bmatrix}$$
$$= [(G \oplus T \oplus C) \quad (A \oplus T \oplus C) \quad (A \oplus T \oplus T)] = [A \quad G \quad A] \qquad [52]$$

The Equation 52 proves that $r^{DNA} = (GAATTC)$ is not a codeword in $c^{DNA}_{(6,3)}$.

If $r^{DNA}$ is codeword other than the transmitted codeword, then *undetected error* occurs. This happens when DNA error pattern, $e^{DNA}$, is also a codeword. We know that in DNA linear block code, the linear combination of two codewords is also a codeword. Thus, if $e^{DNA}$ is a codeword, it transforms $r^{DNA}$ into some other codeword (as $r^{DNA} = c^{DNA} \oplus e^{DNA}$).

### 3.1.5. Error Detection and Error Correction DNA linear Block Code

In this subsection we will discuss the crucial decoding stage of DNA *(n,k)* linear block codes. Equation 50 can be expressed in the following form.

$$\begin{aligned} S^{DNA} &= (S_0^{DNA}, S_1^{DNA} \ldots\ldots S_{n-k-1}^{DNA}) = r^{DNA}(H^{DNA})^T \\ &= (c^{DNA} \oplus e^{DNA})\,(H^{DNA})^T \\ &= c^{DNA}(H^{DNA})^T \oplus e^{DNA}\,(H^{DNA})^T \\ &= (A\,A \ldots.\,A) \oplus e^{DNA}\,(H^{DNA})^T \\ &= e^{DNA}\,(H^{DNA})^T \end{aligned} \qquad [53]$$

(As the DNAX of $(A\,A \ldots.\,A)$ with any other matrices results into the same matrix)

Whenever a *non-A* error syndrome is obtained, the decoder detects that at least one error has been occurred. The decoding can be summarized in following three steps.

- $S^{DNA} = r^{DNA}(H^{DNA})^T$ is calculated.
- The corresponding error pattern ($e^{DNA}$) of $S^{DNA}$ is obtained from the predefined syndrome decoding table.
- The decoder deduces the corresponding error-free codeword as, $c^{DNA} = r^{DNA} \oplus e^{DNA}$.

The steps of error detection and error correction have been explained for $c^{DNA}_{(7,4)}$ in Example 16.

**Example 16:** In this example $c^{DNA}_{(7,4)}$ is considered. First the corresponding syndrome decoding table is calculated. The number of bases in the error pattern is $n$ = 7 and for correctly transmitted codeword the error pattern is *all-A* i.e. ($AAAAAAA$). The bases other than $A$ i.e., $T$ or $G$ or $C$ in any position of the error pattern indicates that error has been occurred. To prepare the syndrome decoding table we need to calculate the corresponding $S^{DNA}$ for each of the 21 possible error patterns as $t^{DNA}$=1 (Equation 48).

If $e^{DNA} = (TAAAAAA)$, the corresponding $S^{DNA}$ can be calculated from Equation 53.

$$S^{DNA} = e^{DNA}\,(H^{DNA})^T = (TAAAAAA)\begin{bmatrix}1 & 0 & 0\\ 0 & 1 & 0\\ 0 & 0 & 1\\ 1 & 1 & 0\\ 0 & 1 & 1\\ 1 & 1 & 1\\ 1 & 0 & 1\end{bmatrix} = (TAA) \quad [54]$$

If $e^{DNA} = (AACAAAA)$, the corresponding $S^{DNA}$ is,

$$S^{DNA} = e^{DNA}\,(H^{DNA})^T = (AACAAAA)\begin{bmatrix}1 & 0 & 0\\ 0 & 1 & 0\\ 0 & 0 & 1\\ 1 & 1 & 0\\ 0 & 1 & 1\\ 1 & 1 & 1\\ 1 & 0 & 1\end{bmatrix} = (AAC) \quad [55]$$

If $e^{DNA} = (AAAAAGA)$, the corresponding $S^{DNA}$ is,

$$S^{DNA} = e^{DNA}\,(H^{DNA})^T = (AAAAAGA)\begin{bmatrix}1 & 0 & 0\\ 0 & 1 & 0\\ 0 & 0 & 1\\ 1 & 1 & 0\\ 0 & 1 & 1\\ 1 & 1 & 1\\ 1 & 0 & 1\end{bmatrix} = (GGG) \quad [56]$$

Thus, calculating $S^{DNA}$ for each of the 21 possible error patterns of DNA (7,4) linear block code the following syndrome decoding table has been prepared (Table 5).

**Table 5:** *Syndrome decoding table for* $c^{DNA}_{(7,4)}$.

| $e^{DNA}$ | $S^{DNA}$ |
|---|---|
| *AAAAAAA* | *AAA* |
| *(T/G/C)AAAAAA* | *(T/G/C)AA* |
| *A(T/G/C)AAAAA* | *A(T/G/C)A* |
| *AA(T/G/C)AAAA* | *AA(T/G/C)* |
| *AAA(T/G/C)AAA* | *(T/G/C)(T/G/C)A* |
| *AAAA(T/G/C)AA* | *A(T/G/C)(T/G/C)* |
| *AAAAA(T/G/C)A* | *(T/G/C)(T/G/C)(T/G/C)* |
| *AAAAAA(T/G/C)* | *(T/G/C)A(T/G/C)* |

Finally, we will explain how errors can be detected and corrected in our proposed DNA linear block code. Let, the received codewords are $r_1^{DNA} = (CGGGTCA)$ and $r_2^{DNA} = (CGTTAAG)$. We must find out the corresponding $c_1^{DNA}$ and $c_2^{DNA}$.

- Now, in the first step of decoding the corresponding syndrome must be calculated.

$$S_1^{DNA} = r_1^{DNA}(H^{DNA})^T = (CGGGTCA)\begin{bmatrix}1 & 0 & 0\\ 0 & 1 & 0\\ 0 & 0 & 1\\ 1 & 1 & 0\\ 0 & 1 & 1\\ 1 & 1 & 1\\ 1 & 0 & 1\end{bmatrix} = (GGA) \quad [57]$$

$$S_2^{DNA} = r_2^{DNA}(H^{DNA})^T = (CGTTAAG)\begin{bmatrix}1 & 0 & 0\\ 0 & 1 & 0\\ 0 & 0 & 1\\ 1 & 1 & 0\\ 0 & 1 & 1\\ 1 & 1 & 1\\ 1 & 0 & 1\end{bmatrix} = (ACC) \quad [58]$$

- From syndrome decoding table (Table 5), the corresponding error patterns, i.e., $e_1^{DNA} = (AAAGAAA)$ and $e_2^{DNA} = (AAAACAA)$, can be selected.
- The corresponding corrected codewords can be produced by,

$$c_1^{DNA} = r_1^{DNA} \oplus e_1^{DNA} = (CGGGTCA) \oplus (AAAGAAA) = (CGG\mathbf{A}TCA) \quad [59]$$

$$c_2^{DNA} = r_2^{DNA} \oplus e_2^{DNA} = (CGTTAAG) \oplus (AAAACAA) = (CGTT\mathbf{C}AG) \quad [60]$$

The corrected base is marked in red in Equations 59 and 60.

In this section we have illustrated the codeword generation, error detection and error correction for mainly DNA (7,4) linear block codes. We have also considered few examples on DNA (6,3) linear block codes. Single error can be detected and corrected in $c_{(7,4)}^{DNA}$ and $c_{(6,3)}^{DNA}$. But the random-error-correcting capability of a DNA linear block code depends on the minimum distance ($d_{min}^{DNA}$) which differs with the length of the codeword and the coding technique. The proposed design strategy of DNA linear block code is also applicable for all the block codes of different sizes.

Our proposed model is different from the existing models of DNA cryptography as we have merged the traditional mathematical and logical techniques of linear block code with the concept of DNA computation. In the paper [20], May et al. used an encoder which maps the genetic code alphabets i.e. *A*, *T*, *G*, *C* into binary bits. On the contrary, in our model, we converted the data into DNA codewords which are consist of only the alphabets of genetic code. Newly presented DNA-based XOR operation (DNAX) is applied for generation, error detection of DNA codewords. DNAX operation has uniqueness and reflexive properties which mimics binary XOR operation.

## 4. Constraints of DNA Codewords Generation

In section 3 we have demonstrated the generation of single stranded DNA codewords by DNA linear block codes. The inter-molecular and intra-molecular hybridizations should be avoided to design an efficient set of DNA codewords. Let, the finite set of DNA alphabets is $\Sigma_{DNA} = \{A, T, G, C\}$. $\Sigma_{DNA}^*$ is the free monoid generated by the set $\Sigma_{DNA}$. If, $C$ is the finite set of DNA codewords, then essentially, $C \subseteq \Sigma_{DNA}^*$. The constraints which can invade the design of competent set of DNA codewords are,

- **Formation of undesirable hairpin structure:** Let, $xy \in C^*$and $x, y \in \Sigma_{DNA}^*$. The reverse complentary sequence of $x$ is denoted by $\overleftarrow{\bar{x}}$. The intra-molecular hybridization of the single stranded DNA sequences leads to the formation of secondary hairpin structure. This can occur in two circumstances; either $\overleftarrow{\bar{y}}$ is the subword of $x$ or $\overleftarrow{\bar{x}}$ is the subword of $y$.
- **Formation undesirable double strands:** Let, $x, y \in C$. If $\overleftarrow{\bar{y}}$ is the subword of $x$, inter-molecular hybridization occurs between the DNA codewords which results in the formation of unwanted partial or full double stranded sequences.
- **Undesirable concatenation of several codewords:** Let, $x, y, z \in C$. The inter-molecular interaction concatenates $x$ and $y$ through the hybridization with $z$, if $\overleftarrow{\bar{z}}$ is the subword of both $x$ and $y$.

### 4.1. Development approach for avoiding constraints

The constraints discussed above can be evaded by following a particular design strategy. To construct the set of DNA codewords in DNA coding theory, the first step is to convert the plain text into DNA sequence following specific encoding methodology. If the encoded DNA sequence only contains *Adenine* ($A$) and *Cytosine* (C), then the set of DNA block codes $C$ over the set $\Sigma_{DNA}$ is $C^+ \subseteq \{A, C\}^+$. Then, for every $x \in C^+$, $\overleftarrow{\bar{x}} \notin C^+$[33]. The developed set of codewords would be unable to form undesirable intra-molecular and inter-molecular interactions.

## 5. Conclusion

The paradigm shift from silicon to carbon is evolving in the domain of coding theory and computation through the amalgamation of biological science and computational science. In this paper we have proposed a novel approach for encryption and decryption using DNA codewords by DNA linear block codes which is the fusion of classical coding theory and DNA computing technology. We have focused to present a systematic methodology which is even capable of detecting and correcting error in coded bases that can occur while transmitting encoded data through biological channels. In broader perspective it can be stated that through DNA coding theory we are expanding the paths towards data compression in the form of DNA strands, error-correcting codes, and DNA cryptography. In this paper we have considered DNA (7,4) and (6,3) linear block codes. But this design strategy is also applicable for all the block codes of different sizes. Higher length and distance of the generated codewords leads to more non-vulnerable encryption and decryption strategy. In future course of research, we have planned to explore the possibility of designing cyclic codes and convolution codes in the field of DNA computing.

## Acknowledgement

The first author acknowledges the financial support received as Research Associate fellowship from Council of Scientific & Industrial Research: Human Resource Development Group (CSIR: HRDG), Government of India.